\documentclass[%
  reprint,
  amsmath,amssymb,
  aps,
  prl,
  superscriptaddress,
  nofootinbib,
  showpacs,
  floatfix,
]{revtex4-2}

\usepackage{graphicx}
\usepackage{bm}
\usepackage{hyperref}
\usepackage{xcolor}

\newcommand{\nuc}[2]{$^{#1}\mathrm{#2}$}

\newcommand{\FA}{\ensuremath{F_{2}^{A}}}
\newcommand{\FD}{\ensuremath{F_{2}^{D}}}
\newcommand{\Rem}{\ensuremath{R_{\mathrm{EMC}}}}

\begin{document}

\title{The Alpha Clustering Model of the EMC Effect}

\author{W.~R.~Armstrong}
\email{warmstrong@anl.gov}
\affiliation{Physics Division, Argonne National Laboratory, Lemont, Illinois 60439, USA}

\date{\today}

\begin{abstract}
For more than forty years the EMC effect, the observation that nuclear structure functions 
in deep-inelastic scattering appear to be modified relative 
to the free-nucleon expectation,  
has resisted a consensus microscopic 
explanation. In this letter, we argue that alpha clustering governs the strength and $A$
dependence of the EMC effect and is responsible for the majority of the
associated partonic modification. We motivate and introduce the alpha clustering model of 
the EMC effect (ACMEE), and with a data driven approach, we demonstrate how 
such a model can reproduce the existing deep-inelastic scattering data on 
nuclei.  We make predictions for other nuclei such as $^{16}$O, $^{20}$Ne, and 
$^{48}$Ca and explore its consequences for a spin-dependent EMC effect and 
global analysis of nuclear parton distributions.

%ACMEE presents a new opportunity for investigating the role of Quantum 
%Chromodynamics (QCD) in the structure of light nuclei by providing a new 
%avenue for understanding QCD confinement beyond meson and baryon structure 
%alone.  Finally, the $\alpha$ clustering model provides hints that there may 
%be distinct regimes of partonic modification, correlated with proton shell 
%closures, we speculate could be related to a growing confinement radius in 
%heavier nuclei.

%Connecting short-ranger correlation (SRC) measurements and the high-momentum 
%tail of single nucleon momentum distributions in nuclei, the $\alpha$ 
%clustering model makes  clear predictions for forth come results and possible 
%future experemnts..

\end{abstract}

\maketitle

%% ---------------------------------------------------------------------------
%\paragraph*{Introduction.---}%
The observation by the European Muon Collaboration (EMC)~\cite{EuropeanMuon:1983wih} that the
per-nucleon deep inelastic scattering (DIS) structure function $\FA/A$ differs
from the free nucleon expectation in the valence quark region is nearly as old 
as QCD itself, and it remains the cleanest
demonstration that the quark distributions of a bound nucleon are not those of
a free one.  Despite four decades of theoretical and experimental effort,
the microscopic origin of the EMC effect has resolved into two broad
families of explanations~\cite{Norton:2003cb,Hen:2016kwk,Malace:2014uea}:
\emph{mean-field} models in which every bound nucleon is modified all of the
time~\cite{Cloet:2006bq,Tsushima:2004st}, and \emph{high-virtuality} models in
which only nucleons in short-range correlated (SRC) pairs are strongly
modified~\cite{Weinstein:2010rt,CLAS:2019vsb}.  The two pictures largely
agree on the observed correlation between the EMC slope and the abundance of
SRC pairs~\cite{Weinstein:2010rt} but disagree sharply on what \emph{causes}
that correlation.

In this Letter we argue that both the EMC effect and the SRC population are
better understood as consequences of a third, more primitive phenomenon:
\emph{$\alpha$-particle clustering} in nuclei. 
The existence of the $\alpha$ particle is as old as the field of nuclear physics itself.
The idea that the $\alpha$ is a privileged degree of freedom in nuclei is 
nothing new. Rutherford, Chadwick and Ellis wrote in the first textbook of nuclear
physics that ``the $\alpha$ particle is of primary importance as a unit of the
structure of nuclei''~\cite{Rutherford1930}. Although written in 1930, 
even considering the discovery of the neutron 2 years later, shell-model, 
and the discovery of quarks and gluons,
this statement about alpha clusters in nuclear structure still rings true 
today. 

First we will give a very brief description of the EMC effect as observed in 
DIS off nuclei. Next we motivate the alpha cluster model of the EMC effect from 
data and as a model consistent with properties of other existing explanations.  
We define ACMEE in its simplest form and present predictions for EMC ratio 
measurements, demonstrating the surprising effectiveness of such a simple 
model. Note, we do not have any model parameters in this treatment, reserving a 
thorough and much more lengthy treatment for a future paper. We conclude with a 
discussion of shortcomings of the present model along with more sophisticated 
model extensions, implications for understanding QCD in nuclei, and new 
directions for future experiments.

In DIS, the Bjorken variable $x = Q^2/2p\cdot q$ represents the total momentum 
fraction of the nucleon carried by the struck parton, where $q$ is the 
four-momentum transfer, $Q^2 = -q^2$, $p$ is the struck nucleon four-momentum.  
The ``EMC region'' is approximately defined as starting in the valence quark 
region where $x$  $> 0.3$  up to about $x$  $\simeq 0.7$ above which Fermi 
motion  and  quasi-elastic scattering contributions appear.
The EMC ratio for a nucleus A is defined as $\Rem^{A}
\equiv (\FA/A)/(\FD/2) \simeq (\sigma_A/A)/(\sigma_{D}/2)$, where $F_2$ is the 
structure function measured in deep inelastic scattering.
The size of the EMC effect is often characterized by its slope in this region,   
$|dR_{\mathrm{EMC}}/dx|$ which is often used to represent the results for each 
nucleus as a single number.  Typically, on the low-$x$ side,  $\Rem^{A}(x=0.3)$ 
is around 1 and drops linearly by roughly  $10-15$\% at $x=0.7$. Often an 
isoscalar correction factor is applied to measurements on nuclei with $Z\ne N$, 
along with Coulomb corrections (see~\cite{Arrington:2021vuu} for details).  For 
the present work, only data without the isoscalar correction factor is 
analyzed, however we retain the Coulomb corrections which only become 
significant for heavy nuclei.

\begin{figure}[htb]
  \includegraphics[width=0.48\textwidth,clip,trim=0mm 5mm 5mm 8mm]{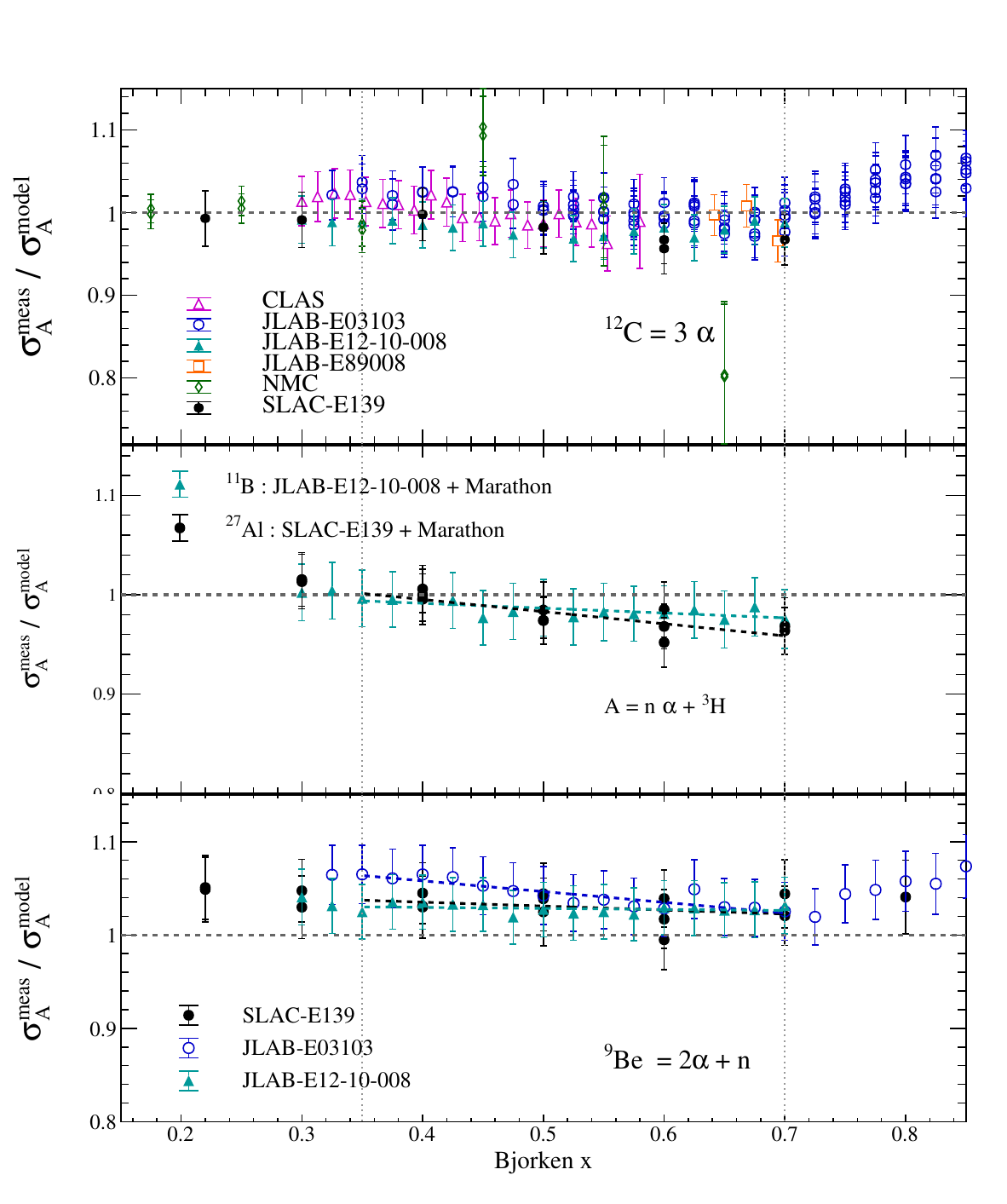}
  \caption{\label{fig:carbon} The  experiment to model ratio for the nuclear 
    cross sections $^{12}$C, $^{11}$B, $^{27}$Al, and $^{9}$Be with data from SLAC 
E139~\cite{Gomez:1993ri}, and JLab E03-103~\cite{Seely:2009gt}, E12-10-008 
\cite{HallC:2022utd,Arrington:2021vuu}, MARATHON \cite{JeffersonLabHallATritium:2024las}, and  CLAS \cite{CLAS:2019vsb}.  }
\end{figure}

One of the first indicators that clustering may have a role in the EMC effect is the observation that
there is very little difference in the EMC ratios for $^4$He to $^{12}$C. This is 
shown in the first panel in Figure~\ref{fig:carbon} where the $^{12}$C cross 
section is modeled as the sum of three $\alpha$s using data from $^4$He, and where 
the $^2$H in the denominators of $\Rem$ cancel leaving the ratio of measurement 
to model prediction. This striking result is often overlooked due to the 
overlays of models and fits (e.g. see \cite{Seely:2009gt}).

The correlation between the EMC slope and the so-called ``plateau'' ($a_2$)
of $\Rem^{A}$ as in inclusive electron scattering experiments for $x > 1$ on nuclei 
motivates the notion that SRCs are somehow responsible for the EMC effect. A 
model based on this idea was developed where the authors claim $n$-$p$ SRC 
pairs are entirely responsible for the EMC effect \cite{CLAS:2019vsb}. They 
find the size of the EMC slope per-proton continues to increase with A while 
the per-neutron size stops at $A=12$.  In this work, we argue that these 
findings,
and short-range correlations themselves, \emph{are downstream consequences of 
the phenomena of $\alpha$ clustering} in nuclei.

The high momentum tail of the single nucleon momentum distribution in nuclei is 
typically associated with short distance nucleon-nucleon configurations, and 
thus SRCs.  Similarly, the EMC slope and local nuclear density 
correlation~\cite{Seely:2009gt} are also naturally explained with the $\alpha$ 
clustering model.  Often attributed to the dominance of $n$-$p$ SRC pairs, the 
isoscalar nature of the EMC effect, if instead originating from the tightly 
bound isoscalar alpha particle, is easily understood.

%That $\alpha$ clustering is more than an emergent modelling artifact was
%established quantitatively by Wiringa, Pieper, Carlson and
%Pandharipande~\cite{Wiringa:2000gb}, who showed with Green's function Monte
%Carlo (GFMC) using the Argonne v18 two-nucleon plus Illinois three-nucleon
%potentials that the ground state of \nuc{8}{Be} exhibits, after rotation of
%each Monte Carlo sample to its principal axis, an unambiguous dinuclear
%two-$\alpha$ density.  The same {\em ab initio} calculations reproduce the
%observed nucleon high-momentum tails that are usually interpreted as SRCs.
%Both features---clustering and the high-momentum tail---emerge from the same
%NN$+$3N Hamiltonian.

Interestingly, in addition to a high momentum nucleons in nuclei, alpha 
clustering emerges naturally from $^8$Be in Quantum Monte-Carlo calculations 
using a modern nuclear potential (AV18) with three-body interactions 
\cite{Wiringa:2000gb}. The same calculations reproduce the
observed nucleon high-momentum tails that are usually interpreted as SRCs.
%Both features---clustering and the high-momentum tail---emerge from the same
%NN$+$3N Hamiltonian.  Furthermore, there is strong correlation between Z and 
%the size of the high momentum tail. 
Thus, it is reasonable to ascribe this 
proton-number dependence with clustering, and subsequently, the observed $n$-$p$ 
dominance of SRCs. 

Turning now to the $\alpha$ clustering model of the EMC effect, we define a 
minimal model in which all of the inputs can be pulled directly from data.  
ACMEE applies to nuclei heavier than $^4$He and builds the nuclei with the following 
ingredients: (1) quasi-free neutrons (no protons) having unmodified partonic 
structure, (2) partial clusters composed of three nucleons which have modified 
partonic structure, and (3) $\alpha$ clusters where the partonic modification 
has saturated. The three-nucleon partial clusters do not have a fully saturated 
partonic modification, however, they do provide an important isospin dependent  
modification. These critical ingredients are provided by data from the MARATHON 
experiment~\cite{JeffersonLabHallATritium:2024las} at JLab which measured the 
EMC ratio for $^{3}$H and $^{3}$He.
In the following data driven analysis we examine the ratio of the experimental 
nuclear cross section to our model predictions.
The nuclear cross sections can then be written as 
$\sigma_A = N_{\alpha} \sigma_{^4\mathrm{He}} + N_{t} \sigma_{^3\mathrm{H}} 
+ N_{h} \sigma_{^3\mathrm{He}} + N_{n} \sigma_{n}$, where the $N_x$ are the 
number of each ingredient with values that follow from the model that nuclei 
preferentially form clusters. For $\alpha$ conjugate nuclei we have $N_{n} = 
N_{h} = N_{t}=0$. For stable light nuclei, as we move up the chart $N_{h}$ and 
$N_{t}$ can be either zero or one, and $N_{n}$ grows with neutron excess.
  
First, we look to reproduce the nuclear cross section for $^{12}$C which is the sum
of three alpha particles, and as already noted, the results agree well with the 
data within their uncertainties as shown in the first panel in 
Figure~\ref{fig:carbon}. This result indicates that for such light nuclei, the 
partonic modification found in $^{4}$He accounts for the observed size of the 
EMC effect. It would be very interesting to compare the model with $^{16}$O and 
$^{20}$Ne nuclear targets where the alpha cluster model predicts similar 
results for these alpha conjugate nuclei.

Next, we use the model to build the iso-vector $N=Z+1$ nuclei $^{11}$B and 
$^{27}$Al as $2\alpha + ^3$H  and $6\alpha + ^3$H, respectively. The results 
are shown in the middle pane of Figure~\ref{fig:carbon} where we make use of 
the MARATHON $^3$H data. The data for these nuclei are remarkably consistent 
with the model predictions which demonstrates the importance of the partial 
clustering configurations contributing \emph{some amount} of partonic 
modification to the tri-nucleon system.
The bottom pane of Figure~\ref{fig:carbon} shows the results for $^9$Be which 
consists of 2 $\alpha$ clusters and a quasi-free neutron.

\begin{figure}[htb]
  \includegraphics[width=0.5\textwidth,clip,trim=8mm 5mm 12mm 15mm]{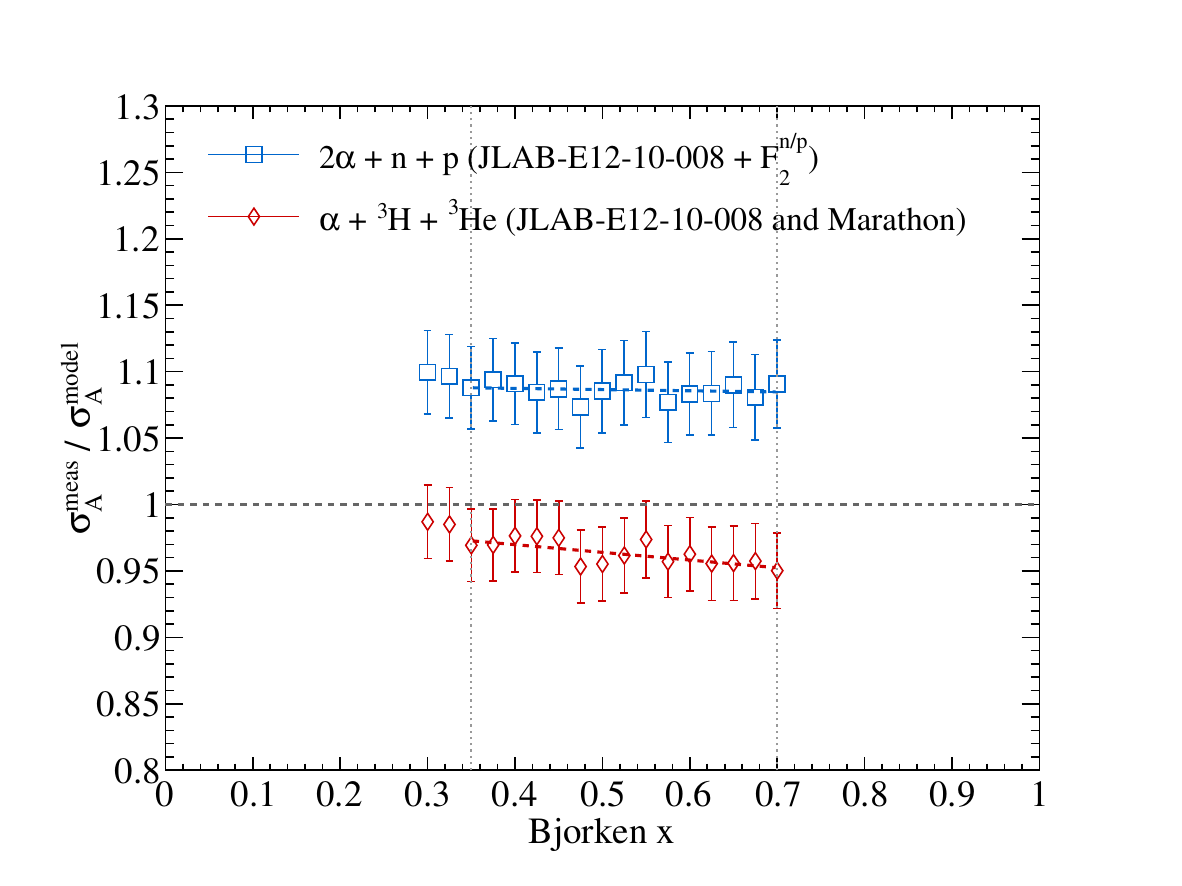}
  \caption{\label{fig:boron10} The ratio of the experimental cross section to 
  the alpha clustering model result  and a model which includes a quasi-free 
proton for $^{10}$B~\cite{HallC:2022utd}.}
\end{figure}

The clustering nature of the EMC effect is further demonstrated with the 
isoscalar nucleus $^{10}$B, which is constructed in the $\alpha$ clustering 
model as the sum of $\alpha +$ $^3$He $+$ $^3$H. Note, the $\alpha$ clustering 
model does not include quasi-free protons as an ingredient. If we had included 
these protons for odd-$Z$ nuclei, $^{10}$B could be modeled as two alphas plus 
a quasi-free $n$-$p$ pair --- in this case the cross section is significantly 
underestimated as shown in Figure~\ref{fig:boron10} where the $\alpha$ 
clustering model prediction nearly agrees with all the data points within their 
uncertainties. Thus, the absence of quasi-free protons in the model represents
the preference for the valence quark distributions in odd-$Z$ nuclei to appear 
as partially clustering, as opposed to forming quasi-free deuteron-like $n$-$p$ 
pairs. 

The few stable odd-$Z$ isoscalar nuclei present very interesting opportunities to 
test predictions of the $\alpha$ clustering model.
Having already looked at $^{10}$B, the others are $^6$Li and $^{14}$N, where 
existing data from NMC \cite{NewMuon:1995cua} and BCDMS \cite{BCDMS:1985dor} 
experiments, respectively, are limited to mostly low-$x$, providing limited 
statistics in the EMC region. If we also consider the spin-dependent EMC effect 
\cite{Cloet:2006bq} these spin-1 nuclei are particularly interesting as we 
discuss below. In ACMEE $^6$Li is constructed as the sum of two partial clusters 
$^3$H+$^3$He. Although this description disagrees with standard shell-model 
description consisting of an $\alpha$ plus deuteron-like $n$-$p$ system, we aim 
to merely describe the valence parton structure of nuclei. Remarkably, this was 
already understood in the early years of QCD where quark potential models of 
nuclei already identified this tension~\cite{Robson:1978vh,Robson:1980hp}.  
Similarly, $^{14}$N is a
4 cluster system (2 partial, 2 fully clustered) and is constructed as 2$\alpha 
+$ $^3$H $+ ^3$He. In the case  of $^{14}$N, this already reflects the 
preference to form a $4\alpha$ cluster system, which shouldn't be too 
surprising considering the doubly magic nucleus $^{16}$O is the fully clustered 
configuration.

If we also consider the spin-dependent or polarized EMC Effect 
(pEMC)~\cite{Cloet:2006bq} in the clustering model, we must  add the following 
ingredients: (1) quasi-free neutrons contribute $g_1^n$ to the spin-dependent 
cross section, (2) the partially clustered systems contribute partially 
modified helicity dependent parton distributions, (3) the spin-zero $\alpha$ 
clusters do not contribute to the spin-dependent cross section, rather their 
presence dilutes the double spin asymmetry. For the partially clustered 
systems, the spin structure function $g_1^{^3\mathrm{He}}$ is well measured, 
but assumed to free of parton-level modification.  On the other hand, 
$g_1^{^3\mathrm{H}}$ remains unmeasured because it would require a polarized 
tritium target. What is more, because the proton carries most of the spin in 
$^3$H, the partonic modification associated with the pEMC effect would 
presumably accompany $g_1^{^3\mathrm{H}} \simeq g_1^p$, making data driven 
model predictions impossible at the moment. 

However, parton-level models of $g_1^{^3\mathrm{H}}$ would be useful for 
$\alpha$ cluster model predictions, unfortunately the pEMC effect mean-field 
calculations are not applicable to $A=3$ nuclei.  Clean measurements of the 
spin structure functions in $A=3$ mirror nuclei would enhance our understanding 
of the polarized EMC effect and provide important data for the spin-dependent 
$\alpha$ clustering model in the same way the MARATHON results did for the 
spin-independent analysis in this letter.  The pEMC Effect in $^{15}$N and 
$^7$Li would also be very interesting for ACMEE predictions since the extra 
neutron completes the $^3$He partial clusters, leaving the $^3$H partial 
cluster to carry all the spin-dependence of the cross section.

\begin{figure}[htb]
  \includegraphics[width=0.50\textwidth,clip,trim=0mm 0mm 10mm 
  15mm]{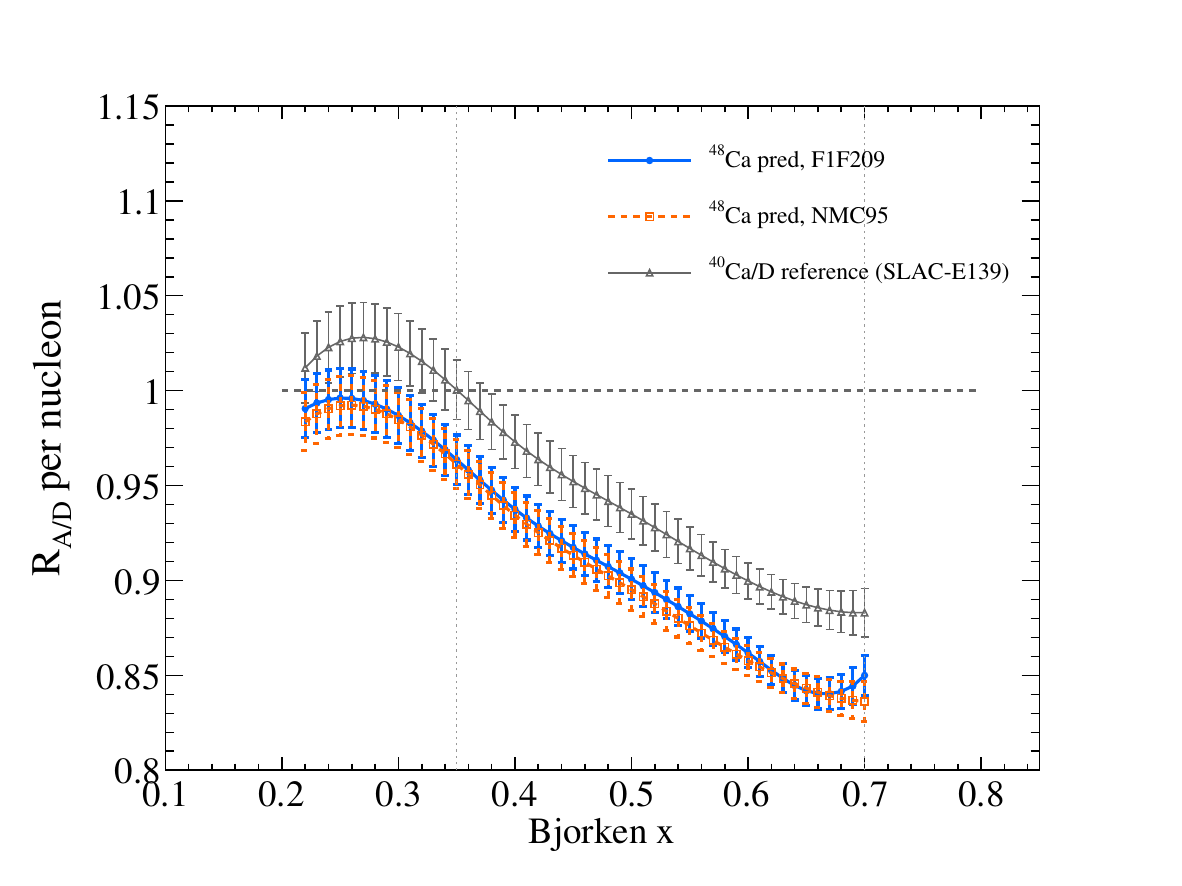}
  \caption{\label{fig:ca48} The ACMEE prediction for the nuclear ratio for 
  $^{48}$Ca which uses the existing data from SLAC-E139~\cite{Gomez:1993ri} and 
the neutron structure function, $F_2^n$, calculated from fits to cross section 
data~\cite{NewMuon:1995zeb,Christy:2007ve}. }
\end{figure}

One  prediction from the $\alpha$ cluster model is the anticipated results for 
the EMC ratio for $^{48}$Ca~\cite{E1210008} as 
$R_{\mathrm{EMC}}^{^{48}\mathrm{Ca}} = (40/48) 
R_{\mathrm{EMC}}^{^{40}\mathrm{Ca}} + (8/48) 2 F_2^n/(F_2^p + F_2^n)$. In ACMEE
the addition of 8 quasi-free neutrons adds the bulk of the EMC offset as shown 
in Figure~\ref{fig:ca48}, effectively shifting the $^{40}$Ca ratio down.  
Within the current uncertainties, the result doesn't significantly vary with 
the structure function model used, where we have used different empirical fits: 
F1F209~\cite{Christy:2007ve} and NMC~\cite{NewMuon:1995zeb}.  A recent 
measurement of proton knockout from \nuc{40}{Ca} and \nuc{48}{Ca} 
\cite{Nguyen:2026ztp}
observed that the abundance of high-momentum short-range pairs depends on the 
shell occupancy of the removed nucleon in a way that a purely 
$NN$-tensor-driven picture of SRCs does not predict.  The
authors summarize the finding as ``nuclear shell structure governs
short-range nucleon pairing.'' Independent of the specific interpretation
offered there, the result is naturally accommodated in the cluster picture:
because the $\alpha$ is the first closed sub-shell of the nuclear shell
model, any organization of nucleons into $\alpha$-like sub-shells is by
construction a shell-model statement.  In a nucleus like \nuc{48}{Ca}, where
the extra neutrons occupy an $f_{7/2}$ shell that lies outside the natural
$\alpha$-clustering support of the $sd$-shell core, they do not participate
in $\alpha$-generated short-range correlations at the same rate as the core
nucleons of \nuc{40}{Ca}. This is evidenced in a recent measurements of $x>1$
inclusive electron elastic scattering~\cite{Zhang:2025nst} where the reported 
$a_2$ for $^{48}$Ca is lower than values for $^{40}$Ca and $^{12}$C.

%In an SRC-only picture this shell dependence has
%to be added by hand; in a cluster picture it is built in.
%Further evidence appears in recent results looking at SRCs in (e,e'p) 
%coincidence measurements \cite{} and inclusive scattering at $x>1$ on 
%$^{48}$Ca where the proton number seems to be the determining factor.  In the 
%latter measurement, the $a_2$ values hardly appear as plateaus but rather 
%appear to be an inflection point and the 48Ca value appears lower than 40.

Clearly more precise data from a broader range of nuclei could help validate 
this model and the clustering origin of the EMC effect.
In particular, precision measurements on $^4$He,  $^{12}$C, $^{16}$O, and $^{20}$Ne with 
improved systematic uncertainties would go a long way in this direction. 
For example, using a high-pressure gaseous straw target would mitigate 
systematic uncertainties associated with different target geometries 
(e.g. thin foil vs extended cryogenic liquid) as is proposed at JLab~\cite{Armstrong:2026yyz}.

The $\alpha$ clustering model places all the partonic modification in $A\le4$ clustering configurations,
thus leading to a significant departure from the typical monotonic $A$ dependence assumed in global
nuclear PDF analyses~\cite{nCTEQ:2026ewr,Eskola:2021nhw}.
%Outlook for the $\alpha$ cluster model of the EMC effect:
%The extraction of the nuclear parton distributions from global analysis 
%typically assume a nuclear dependence as function of A~\cite{nCTEQ:2026ewr}.
In the $\alpha$ clustering model there are only two types of modified valence parton 
distributions, those associated with a partially clustered configuration ($^3$H or $^3$He) 
or a fully clustered $\alpha$ particle. The partially modified configuration $^3$H 
provides a particularly interesting feature in the form of the isovector EMC 
effect \cite{Cloet:2009qs,Cocuzza:2021rfn}, especially useful for modeling 
heavier nuclei. As neutron excess grows, $^3$H partial clusters are an 
attractive mechanism for generating a neutron skin in connection with the 
isovector EMC effect.

The lack of a quasi-free proton in ACMEE causes tension with the shell model picture
for light nuclei and is an area open for further theory and experimental investigation. 
Again, it should be emphasized that only the valence parton distributions in DIS are 
considered in the model. However, as new experiments probing the partonic structure of
$^4$He are completed or underway~\cite{CLAS:2017udk,Armstrong:2017wfw,alert2}, it begs the question: is the partonic structure of a 
bound $\alpha$ particle modified in nuclei~\cite{Armstrong:2026yyz}? An example more to the point,
the QCD inspired hexadiquark model~\cite{West:2020rlk} identifies a maximally 
hidden color configuration of the 12 quark system which may form the core of the $\alpha$
clusters. Such a substructure may provide the Pauli blocking necessary for 
quark potential models~\cite{Robson:1978vh}, or perhaps more naturally supply a source  
for Pauli-forbidden states in $\alpha$ cluster models generating an 
apparent ``supersolidity'' in nuceli~\cite{Ohkubo:2019one,Ohkubo:2022fgo}. 
Each of these very different models imply the existence of non-nucleonic structure of nuclei
where the alpha particle is playing a central role.

%In summary, we propose a new model for the EMC effect based on $\alpha$ clustering
%which consistently reproduces the nuclear data for the EMC effect. 
In  summary,  the minimal form of ACMEE is a parameter-free, data driven construction: nuclei are
assembled from quasi-free neutrons with unmodified partonic structure, partially
clustered tri-nucleon systems whose modification is fixed by 
\nuc{3}{H} and \nuc{3}{He} data, and fully clustered $\alpha$ particles whose partonic structure is
saturated in \nuc{4}{He}. With all inputs taken directly from measurement, the model
reproduces the EMC ratios of nuclei within their uncertainties, 
indicating that the majority of the EMC effect originates in
the bound $\alpha$, while the residual $A$ dependence follows from how, and how
much, each nucleus organizes its nucleons into $\alpha$ and partial clusters. In
this picture, the observed correlation between the EMC slope and the SRC scaling
factor $a_2$, the $n$-$p$ dominance of SRC pairs, and the recent finding that nuclear
shell structure governs short-range nucleon pairing are all natural consequences of
$\alpha$ clustering rather than independent phenomena. The model provides
falsifiable predictions for the $\alpha$-conjugate nuclei \nuc{16}{O} and
\nuc{20}{Ne} and for the forthcoming \nuc{48}{Ca} measurement, and it can be
extended to the spin-dependent EMC effect. These tests will determine whether the $\alpha$
particle is the primary carrier of partonic modification in nuclei which has  
direct consequences for the global extraction of nuclear parton distributions and 
the role of QCD in nuclear structure.

\begin{acknowledgments}
The author thanks I.~C.~Clo\"et, R.~B.~Wiringa, and D.~F.~Geesaman for
discussions on cluster structure, ab~initio wave functions, and the
history of the EMC effect. We also thank Sangbaek Lee for his feedback and 
careful reading of this manuscript. This work is supported by the U.S.~Department
of Energy, Office of Science, Office of Nuclear Physics, under contract
DE-AC02-06CH11357.
\end{acknowledgments}

%% ---------------------------------------------------------------------------
%% References
%% ---------------------------------------------------------------------------
%% NOTE: This draft cites entries expected to live in ../2026_tomography_workshop/bibliography.bib
%%       plus a small number of additional standard EMC/SRC references that must
%%       still be added there.  Missing keys are flagged below.
%%

\bibliographystyle{apsrev4-1}
\bibliography{refs}% Produces the bibliography via BibTeX.

\end{document}